\documentclass[twocolumn,aps,superscriptaddress,showpacs,nofootinbib,floatfix]{revtex4}
\usepackage{epsfig,bm,feynmf}
\usepackage{graphicx}
\usepackage{amsmath}
\usepackage{dcolumn}

\usepackage{graphicx}
\usepackage[usenames,dvipsnames,svgnames,table]{xcolor}

\usepackage{tikz-feynman,contour}
\tikzfeynmanset{compat=1.0.0}

\begin{document}


\title{Charmonium production in a thermalizing heat bath}


\author{Taesoo Song}\email{t.song@gsi.de}
\affiliation{GSI Helmholtzzentrum f\"{u}r Schwerionenforschung GmbH, Planckstrasse 1, 64291 Darmstadt, Germany}

\author{Joerg Aichelin}\email{aichelin@subatech.in2p3.fr}
\affiliation{SUBATECH UMR 6457 (IMT Atlantique, Universit\'{e} de Nantes, IN2P3/CNRS), 4 Rue Alfred Kastler, F-44307 Nantes, France}
\affiliation{Frankfurt Institute for Advanced Studies, Ruth-Moufang-Strasse 1, 60438 Frankfurt am Main, Germany}

\author{Elena Bratkovskaya}\email{E.Bratkovskaya@gsi.de}
\affiliation{GSI Helmholtzzentrum f\"{u}r Schwerionenforschung GmbH, Planckstrasse 1, 64291 Darmstadt, Germany}
\affiliation{Institute for Theoretical Physics, Johann Wolfgang Goethe Universit\"{a}t, Frankfurt am Main, Germany}
\affiliation{Helmholtz Research Academy Hessen for FAIR (HFHF),GSI Helmholtz Center for Heavy Ion Physics. Campus Frankfurt, 60438 Frankfurt, Germany}


\begin{abstract}
Using the Remler formalism for the creation of composed particles, we study chamonium production both in thermalized and thermalizing boxes, which contain charm and anticharm quarks.  The thermalizing box studies include the lowering of the box  temperature, the spatial diffusion of charm and anticharm quarks, which are initially confined in the central region, as well as the combination of both, which imitates heavy-ion collisions.
Comparing numerical and analytical results we demonstrate that the rate of the original Remler formalism has to be supplemented by two rates  to obtain, for $t\to \infty$, results, which are consistent with the statistical model predictions:  i) a rate, which takes into account the temperature dependence of the Wigner density of the quarkonium during the expansion and,  in the case that a heavy quark potential is not implemented in the Monte Carlo approach, ii) a rate which comes from the change of the relative distance between the heavy quark and antiquark.
These results provide the basis for future applications of the Remler formalism to heavy-ion collisions.
\end{abstract}



\maketitle

\section{Introduction}

Quarkonium is a bound state of a heavy quark and its antiquark.
Since the object is flavorless, it is often called hidden heavy flavor meson.
Several decades ago Matsui and Satz suggested the suppression of quarkonium production in heavy-ion collisions as a signature for the formation of  a quark-gluon plasma (QGP), because the color screening, which exists only in a deconfined phase, prevents the heavy quark pair from forming a bound state~\cite{Matsui:1986dk}. 

Quarkonium suppression was indeed later observed at the Super Proton Synchrotron (SPS) and the Relativistic heavy-ion Collider (RHIC)~\cite{NA50:2004sgj,PHENIX:2006gsi}. As the collision energy increases, however, more and more heavy quarks are produced and the possibility of regeneration of quarkonia emerges also because  the c and $\bar c$ from different primary vertices may form a quarkonium. In fact, the nuclear modification factor of $J/\psi$, which measures the normalized ratio of $J/\psi$ produced in heavy-ion reactions as compared to pp collisions,  is larger at the Large Hadron Collider (LHC) than at RHIC and larger in mid-rapidity than in forward and backward rapidities, though the temperature is higher at LHC and in the mid-rapidity region ~\cite{Manceau:2012ka}. 
Therefore the study of quarkonium production in heavy-ion collisions is not simply focused on the dissociation in a hot dense nuclear matter but also on regeneration in a QGP~\cite{Grandchamp:2002wp,Yan:2006ve,Linnyk:2008hp,Song:2011xi} or when the QGP hadronizes.

In the seventies Remler devised a formalism to study the production of composite particles in heavy-ion collisions by using the Wigner representation of density operators~\cite{Remler:1975re,Remler:1975fm,Remler:1981du}, which was successfully applied to deuteron production in heavy-ion collisions~\cite{Remler:1975re,Gyulassy:1982pe}.

Recently an attempt has been made to use the Remler formalism to study quarkonium production in heavy-ion collisions~\cite{Villar:2022sbv,Villar:2021loy} as well as in p+p collisions~\cite{Song:2017phm}.
A distinguished feature of this approach is that the temperature and therefore in a QGP time dependence  of the Wigner density of the  quarkonium is taken into account. 
Such a dependence is predicted by lattice gauge calculations \cite{Lafferty:2019jpr}, which show that below the dissociation temperature (above which a $J/\psi$ is not stable anymore) the root-mean-square (rms) radius of a $J/\psi$ depends on the temperature. 

In this study we apply the Remler formalism for quarkonium production in a thermal box by using Monte-Carlo methods.  Contrary to simulations of heavy-ion collisions, box simulations have the advantages that everything is controllable and analytic solutions are available.  The equilibration of the quarkonia in the box is achieved by scattering with virtual particles at a given temperature.  The asymptotic distribution can then be compared with statistical model predicitions.  The statistical model, which has successfully described the particle production in heavy-ion collisions, provides reliable solutions for the quarkonium production~\cite{Andronic:2005yp,Andronic:2006ky}.

We first briefly review the Remler formalism in Sec.~\ref{Remler} and present  the solutions for quarkonium production in a thermalized box in Sec.~\ref{coalescence}.
In the next section we carry out box simulations for four different initial conditions and 
present our results.
In Sec.~\ref{diff-sec} the necessity of an additional term, which is responsible for spatial diffusion, is explained.
Finally, a summary is given in Sec.~\ref{summary}.

\section{Remler formalism}\label{Remler}

In the Remler formalism the multiplicity of a quarkonium state  $\Phi$, $P_\Phi$, in a $N$-body system, composed of heavy (anti)quarks, is asymptotically (for $t\to\infty$) obtained by
\begin{eqnarray}
P_\Phi(t\to \infty) ={\rm Tr}[\rho_\Phi \rho_N(t\to\infty)],
\label{density}
\end{eqnarray}
where $\rho_\Phi$ (which is assumed to be time independent) and $\rho_N$ are, respectively, the density operator of the quarkonium and of the $N$ heavy (anti)quarks.

In practice we cannot solve the time evolution of the quantal  N-body density matrix without approximations.
In the past it turned out that very satisfying results are obtained, if one does not study the time evolution of the density matrix itself but that of the Wigner density distribution, the Fourier transform of the density matrix, and approximates, guided by the mathematical work on the solution of the quantal Vlasov equation,  the quantal Wigner density distribution by an ensemble of classical phase space distributions of point like particles.
Averaging over this ensemble, one can calculate observables, which agree remarkable well with the experimental results.  This procedure is the background of the Boltzmann-Uehlung -Uhlenbeck (BUU)  (cf.~\cite{Aichelin:1985zz,Cassing:2021fkc}) and the Vlasov-Uehling-Uhlenbeck (VUU) \cite{Kruse:1985pg} approach, which are widely used to describe the results of heavy-ion collisions with center of mass energies between few GeV and several TeV
(cf. \cite{Petersen:2018jag,Linnyk:2008hp,Bleicher:2022kcu}).

In these approaches the particles scatter and move on curved trajectories due to a
mean potential, created by the fellow particles.  If this potential is absent or neglected one talks about a cascade approach.  If heavy (anti)quarks  are described in the cascade mode, where only scattering and free streaming but no potential interaction is present, they will have diverging trajectories. Therefore, even if  the heavy (anti)quarks are initially confined to a space region, asymptotically we will find 
\begin{eqnarray}
\lim_{t\rightarrow 0} P_\Phi=0,
\end{eqnarray} 
because two and more body correlations are lost in this approach. To overcome this problem one introduces a rate 
\begin{eqnarray}
&&\Gamma=\frac{d P_\Phi}{dt}={\rm Tr}\bigg(\frac{d\rho_\Phi}{dt} \rho_N\bigg)+{\rm Tr}\bigg(\rho_\Phi \frac{d\rho_N}{dt}\bigg) \label{gamma}\\
&&={\rm Tr}\bigg(\frac{d\rho_\Phi}{dt} \rho_N\bigg)-i{\rm Tr}\bigg(\rho_\Phi [H,\rho_N]\bigg)\equiv \Gamma_{\rm local}+\Gamma_{\rm coll}. \nonumber
\end{eqnarray}
The Hamiltonian is decomposed into
\begin{eqnarray}
H&=&\sum_i K_i+\sum_{i<j} V_{ij} \label{Hamilt}\\
&=&H_{1,2}+H_{N-2}+\sum_{i\ge 3}(V_{1i}+V_{2i}), \nonumber
\end{eqnarray}
where $K_i$ and $V_{ij}$ are, respectively, the  kinetic and interaction terms and
\begin{eqnarray}
H_{1,2}&=&K_1+K_2+V_{12},\nonumber\\
H_{N-2}&=&\sum_{i\ge 3}K_i+\sum_{i>j\ge 3}V_{ij}.
\label{hamiltonian}
\end{eqnarray}
Using the cyclic property of traces,
\begin{eqnarray}
\Gamma_{\rm coll}=-i{\rm Tr}\bigg(\rho_\Phi [H,\rho_N]\bigg)=i{\rm Tr}\bigg(\rho_N [H,\rho_\Phi]\bigg),
\end{eqnarray}
and supposing for simplicity  that $\Phi$ contains the particles 1 and  2 we obtain
\begin{eqnarray}
[H_{N-2},\rho_\Phi]=0,~~~~~[H_{1,2},\rho_\Phi]=0,
\label{h12}
\end{eqnarray}
because $H_{N-2}$ does not affect $\rho_\Phi$ and $\rho_\Phi$ is a eigenfunction of $H_{1,2}$. 
The collision term in Eq.~(\ref{gamma}) is then simplified to~\cite{Gyulassy:1982pe,Villar:2022sbv}
\begin{eqnarray}
\Gamma_{\rm coll}=-i\sum_{i\ge 3}{\rm Tr}\bigg(\rho_\Phi [V_{1i}+V_{2i},\rho_N]\bigg).
\label{potential}
\end{eqnarray}

For a S-state the density operator of  the quarkonium, expressed in Wigner representation, is approximated by
\begin{eqnarray}
\rho_\Phi\rightarrow W_S(r,p)=8\exp\bigg[-\frac{r^2}{\sigma^2}-\sigma^2 p^2\bigg],
\label{1sWig}
\end{eqnarray}
where $r$ and $p$ are, respectively, relative distance and relative momentum in the center-of-mass frame. The width $\sigma$ is given by the rms radius of the quarkonium.

The classical phase space distribution of point like particles can be expressed as
\begin{eqnarray}
\rho_N\approx \prod_{i=1}^N h^{3N}\delta(r_i-r_i^*(t))\delta(p_i-p_i^*(t))
\label{deltas}
\end{eqnarray}
where $r_i^*(t)$ and $p_i^*(t)$ is the phase space trajectory of particle $i$. The 
time derivative of the density matrix is then given by
\begin{eqnarray}
&&\frac{d\rho_N}{dt}\approx \sum_i v_i\cdot \nabla_r \rho_N   \label{impulse}\\
 &&+\sum_{i>j}\sum_\nu \delta(t-t_{ij}(\nu))\{\rho_N(t+\varepsilon)-\rho_N(t-\varepsilon)\} \nonumber
\end{eqnarray}
where $t_{ij}(\nu)$ is the time for the $\nu$ th collision between  the particles $i$ and $j$. 
The first term implies free streaming between instant scatterings, which are described  by the second term.  The second term in Eq.~(\ref{impulse}) is equivalent to Eq.~(\ref{potential}).
Therefore,
\begin{eqnarray}
\Gamma_{\rm coll}(t)&\approx& \sum_{i=1,2}\sum_{j\ge 3}\sum_\nu \delta(t-t_{ij}(\nu)) \label{gam-old} \\
&\times& \int d^3r_1d^3p_1 ... d^3r_N d^3p_N (2\pi)^{3N}\nonumber\\
&\times& \rho_\Phi(r_1,p_1;r_2,p_2)\{\rho_N(t+\varepsilon)-\rho_N(t-\varepsilon)\},\nonumber
\end{eqnarray}
and we obtain for the multiplicity at time t'
\begin{eqnarray}
P_\phi(t^\prime)-P_\phi(0) \approx \int^{t^\prime}_0  dt \{\Gamma_{\rm local}(t)+ \Gamma_{\rm coll}(t)\}.
\label{joerg}
\end{eqnarray}
We note that $\Gamma_{\rm local}(t)$ contributes only if $\sigma$ in Eq.~(\ref{1sWig}) changes with time~\cite{Villar:2022sbv}.

\section{quarkonium in thermal box}\label{coalescence}

The multiplicity of a quarkonium $\Phi$ can be obtained in the coalescence approach by projecting the phase space distribution of  the charmed quarks onto the Wigner density of the $\Phi$ state~\cite{Song:2016lfv,Song:2021mvc}
\begin{eqnarray}
N_{\Phi}=\frac{d}{d_1d_2}\int\frac{d^3p_1 d^3p_2 d^3r_1 d^3r_2}{(2\pi)^6} \label{s-wave} \\
\times f_Q(r_1,p_1) f_{\bar{Q}}(r_2,p_2)W_S(r,p)\nonumber\\
=\frac{d}{d_1d_2}\int\frac{d^3P d^3p d^3R d^3r}{(2\pi)^6}\nonumber\\
\times f_Q(r_1,p_1) f_{\bar{Q}}(r_2,p_2)W_S(r,p),\nonumber
\end{eqnarray}
where $ f_Q(r_1,p_1)$ and $f_{\bar{Q}}(r_2,p_2)$ are, respectively, the heavy quark (Q) and heavy antiquark ($\bar Q$) distribution functions, $r=r_1-r_2$, $R=(r_1+r_2)/2$, $p=(p_1-p_2)/2$ and $P=p_1+p_2$, and $d_1$, $d_2$ and $d$ is, respectively, the  spin-color degeneracy of $Q$, $\bar{Q}$ and quarkonium.
$r$ and $p$ in the Wigner functions are the distance and the relative momentum of the heavy quark pair in their center-of-mass frame. Since we are studying heavy quarks with a mass much larger than the temperature, we can safely use the Galilean transformation instead of the Lorentz transformation.

We assume an uniform distribution of $Q$ and $\bar{Q}$ in space,
\begin{eqnarray}
(2\pi)^3 \frac{dN_{S}}{Vd^3P}=\frac{d}{\pi^3 d_1d_2}\int  d^3p d^3r f_Q(p_1) f_{\bar{Q}}(p_2)e^{-\frac{r^2}{\sigma^2}-\sigma^2 p^2}\nonumber\\
=\bigg(\frac{\sigma^2}{\pi}\bigg)^{3/2}\frac{d}{d_1d_2}\int  d^3p f_Q(p_1) f_{\bar{Q}}(p_2) e^{-\sigma^2 p^2},~~~
\label{1s-full}
\end{eqnarray}
because
\begin{eqnarray}
\int d^3r e^{-r^2/\sigma^2}
=2\pi\sigma^{3}\Gamma(3/2)=(\pi\sigma^2)^{3/2}.
\label{cal1}
\end{eqnarray}

In the nonrelativistic  (or heavy quark) limit one can take a Boltzmann distribution for $f_Q(p_1)$ and $f_{\bar{Q}}(p_2)$ 
\begin{eqnarray}
\frac{1}{d_1d_2}f_Q(p_1) f_{\bar{Q}}(p_2) e^{-\sigma^2 p^2}\approx
e^{-E_1/T-E_2/T-\sigma^2 p^2}\nonumber\\
\approx
\exp\bigg[-\bigg(2m_Q+\frac{p_1^2}{2m_Q}+\frac{p_2^2}{2m_Q}\bigg)\bigg/ T-\sigma^2 p^2\bigg]\nonumber\\
=\exp\bigg[-\bigg(2m_Q+\frac{P^2/2+2p^2}{2m_Q}\bigg)\bigg/T-\sigma^2 p^2\bigg].
\label{NR-1s}
\end{eqnarray}
Substituting Eq.~(\ref{NR-1s}) into Eq.~(\ref{1s-full}) we obtain
\begin{eqnarray}
(2\pi)^3 \frac{dN_{S}}{Vd^3P}=d\exp\bigg[-\bigg(M+\frac{P^2}{2M}\bigg)\bigg/T\bigg]\nonumber\\
\times   \bigg(\frac{\sigma^2}{\pi}\bigg)^{3/2}\int  d^3p  e^{-\{\sigma^2 +1/(m_QT)\}p^2}\nonumber\\
=d~ e^{-E/T}\bigg(\frac{\sigma^2}{\sigma^2 +1/(m_QT)}\bigg)^{3/2}
\label{approx-new}
\end{eqnarray}
where $M=2m_Q$.
Assuming $m_Q \gg 1/(\sigma^2 T)$ we find, 
\begin{eqnarray}
(2\pi)^3 \frac{dN_{S}}{Vd^3P}
\approx d~ e^{-E/T}.
\label{approx-s}
\end{eqnarray}
 Since $\sigma^2=8/3 \langle r^2 \rangle$ with $r$ being the quarkonium radius~\cite{Song:2016lfv,Song:2021mvc}, 
the assumption is justified if
\begin{eqnarray}
m_Q \gg \frac{3}{8 \langle r^2 \rangle T}.
\end{eqnarray}
It is hence valid even close to the critical temperature $T_c$, supposing $\sqrt{\langle r^2 \rangle} \sim$~0.5~fm.
We note that from Eq.~(\ref{approx-new})  that the charmonium abundance at $T=200$ MeV is about 77 \% of  the statistical model abundance for $m_Q=1.5$ GeV.

Therefore we expect that in a heat bath the statistical model and the coalescence approach yield a similar multiplicity. Strictly speaking, there must be an attractive force between the heavy quark and the heavy antiquark to form a quarkonium and the mass of the quarkonium is therefore less than twice the heavy quark mass. In this study, however, we assume that the binding energy is small and therefore the $J/\psi$ is weakly bound in a QGP~\cite{Lee:2013dca,Gubler:2020hft,Song:2020kka}. The inclusion of heavy quark potential in the Remler formalism \cite{Villar:2022sbv} will be discussed in a future publication.

The coalescence model of Eq.~(\ref{s-wave}) is closely related to Eq.~(\ref{density})~\cite{Song:2016lfv}, since
\begin{eqnarray}
P_\Phi={\rm Tr}[\rho_\Phi \rho_N]={\rm Tr}\bigg(|\Phi\rangle \langle \Phi|\nonumber\\
\times |Q_1\bar{Q}_2...Q_{N-1}\bar{Q}_N\rangle_i \rho_{ij} \langle Q_1\bar{Q}_2...Q_{N-1}\bar{Q}_N|_j  \bigg)\nonumber\\
=\bigg|\langle \Phi| Q_1\bar{Q}_2...Q_{N-1}\bar{Q}_N\rangle_i \bigg|^2 \rho_{ii}, 
\end{eqnarray}
where $\rho_{ij}$ is the density matrix element of the $N-$body density matrix $\rho_N$ and the eigenstates  $|\Phi\rangle$ are taken for a basis of the $c\bar c$ states. 

The Remler formalism starts from the same equation~(\ref{density}). Therefore we can test this formula in a thermal box, which is completely controllable and for which explicit and analytical solutions of Eq.~(\ref{approx-s}) can be obtained.  That the numerical realization of the Remler formalism gives the correct result in a box  is a prerequisite for its application in numerical simulation of heavy-ion collisions.

\section{box simulations}

In this section the Remler formalism is tested  in a box in which the heavy (anti)quarks are in thermal equilibrium as well as for three scenarios in which their initial momentum distribution and/or their  initial spatial distribution does not correspond to the equilibrium distribution.
\subsection{static thermal box}\label{sim-box}

We prepare a box of size $100^3~{\rm fm^3}$ in which we place charm quarks and charm antiquarks in thermal equilibrium at $T=$ 200 MeV. Charm (anti)quarks scatter off 
artificial partons which have a thermal momentum of $T=$ 200 MeV and are not affected by the scattering. The interaction rate is fixed to 1.0 c/fm. To remove boundary effects, we extended the box by 3 fm in each direction but for the analysis we exclude the extended volume. 

\begin{figure}[h!]
\centerline{
\includegraphics[width=9 cm]{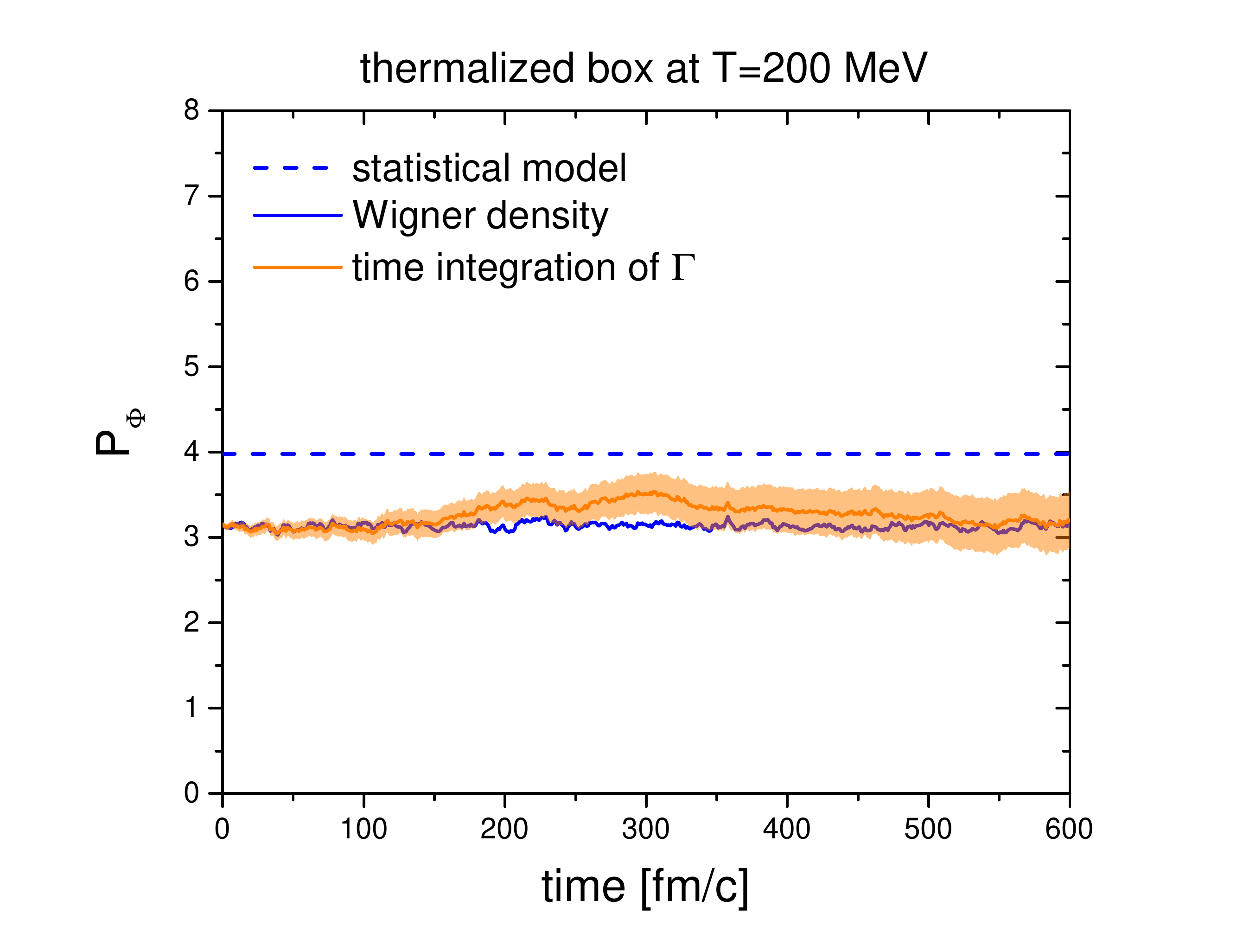}}
\caption{(Color online) Charmonium multiplicity as a function of time in a box of $100^3 ~{\rm fm^3}$ at $T=$ 200 MeV calculated in  the statistical model, from Eq.~(\ref{s-wave}) and from Eq.~(\ref{joerg}). Charm quark mass and charmonium radius are taken as 1.5 GeV and 0.5 fm, respectively.}
\label{box}
\end{figure}

Figure~\ref{box} compares the charmonium multiplicity as a function of time from statistical model calculations, assuming  that the quarkonium mass is twice the heavy quark mass, as in Eq.~(\ref{approx-s}), from Eq.~(\ref{s-wave}) and from Eq.~(\ref{joerg}), which are named in the figure "statistical model", "Wigner density" and "time integration of $\Gamma$", respectively.  The charm quark mass is assumed to be 1.5 GeV and and we take a charmonium radius of 0.5 fm. The results are an ensemble average of 200 events. The colored band indicates the statistical error.
One can see a slight difference between the statistical model and the Wigner projection, which originates from the approximations in Eqs.~(\ref{NR-1s}) and (\ref{approx-s}).

One can see that the time integration of $\Gamma_{\rm coll}$ does not deviate from the dashed and solid blue lines within the statistical error. The reason we will explain now.  

We define the Wigner projection at $t$,
\begin{eqnarray}
\mathcal{W}(r_1^*,  r_2^*, p_1^*, p_2^*; t)\equiv \sum_{i=1,2}\sum_{j\ge 3} \int d^3r_1d^3p_1 ... d^3r_N d^3p_N  \nonumber\\
\times (2\pi)^{3N}\rho_\Phi(r_1,p_1;r_2,p_2)\rho_N(t),~~~
\end{eqnarray}
where $r_1^*,  r_2^*, p_1^*, p_2^*$ 
characterize
trajectories of (anti)cham quarks in Eq.~(\ref{deltas}) projected to quarkonium state, see Eq.~(\ref{deltas}).

Since the projection is carried out in a homogeneous box, one can separate the spatial dependence, which will be a constant in time, such that
\begin{eqnarray}
\mathcal{W}(r_1^*,  r_2^*, p_1^*, p_2^*; t)=\mathcal{W}_p(p_1^*, p_2^*; t)\mathcal{W}_r.
\end{eqnarray}

Then Eq.~(\ref{joerg}) is expressed as
\begin{eqnarray}
P_\phi(t^\prime) = \mathcal{W}_p(p_1^*, p_2^*; 0)\mathcal{W}_r\nonumber\\
+\mathcal{W}_p(p_1^*, p_2^*; t_1+\varepsilon)\mathcal{W}_r- \mathcal{W}_p(p_1^*, p_2^*; t_1-\varepsilon)\mathcal{W}_r\nonumber\\
+\mathcal{W}_p(p_1^*, p_2^*; t_2+\varepsilon)\mathcal{W}_r- \mathcal{W}_p(p_1^*, p_2^*; t_2-\varepsilon)\mathcal{W}_r\nonumber\\
...\nonumber\\
+\mathcal{W}_p(p_1^*, p_2^*; t^\prime+\varepsilon)\mathcal{W}_r- \mathcal{W}_p(p_1^*, p_2^*; t^\prime-\varepsilon)\mathcal{W}_r,
\label{boxi}
\end{eqnarray}
where $t=0$ is the initial projection time where only the  production term appears, which corresponds to $P_\phi(0)$ in Eq.~(\ref{gam-old}). We number the scatterings of heavy quarks or antiquarks by "$i$".  $t_i$ is the time of the $i$'th scattering  in which either the charm quark or the  anticharm quark is involved and both, production and destruction term, appear. We note that $\mathcal{W}_p(p_1^*, p_2^*; t)$ in the above equation means $\mathcal{W}_p(p_1^*(t), p_2^*(t))$.

Since there is no scattering of charm quarks with a light quark between $t=0$ and $t=t_1-\varepsilon$, 
\begin{eqnarray}
\mathcal{W}_p(p_1^*, p_2^*; 0)=\mathcal{W}_p(p_1^*, p_2^*; t_1-\varepsilon).
\label{box1}
\end{eqnarray}
The same applies to each following time interval between collisions:
\begin{eqnarray}
\mathcal{W}_p(p_1^*, p_2^*; t_i+\varepsilon)=\mathcal{W}_p(p_1^*, p_2^*; t_{i+1}-\varepsilon).
\label{box2}
\end{eqnarray}
Thus, most of the terms cancel and only the  Wigner projection at the end of the last collision remains:
\begin{eqnarray}
P_\phi(t^\prime) = \mathcal{W}_p(p_1^*, p_2^*; t^\prime+\varepsilon)\mathcal{W}_r,
\label{boxf}
\end{eqnarray}
where $p_1^*$ and $p_2^*$ are still the thermal momenta of the charm quark and anticharm quark. The time integration of $\Gamma_{\rm coll}$ fluctuates around the Wigner projection for a system in  thermal equilibrium at T=200 MeV.

One can see that the statistical error in Fig.~\ref{box} increases with time.
The reason is as following: Whenever scattering happens, a new Wigner projection is added and the old Wigner projection is subtracted as in Eq.~(\ref{boxi}). 
Since the box is already in thermal equilibrium both the addition and the subtraction are random fluctuations. As a result, like for all random walks, some events deviate far away from the thermal average as time passes. That is why the statistical error increases with time, while the average value stays near the thermal equilibrium.

\subsection{cooling down of (anti)charm quarks}

Now we apply the Remler formalism to charm and anticharm quarks in a box of the same size in which the initial temperature of (anti)charm quarks is 400 MeV, but their number is same as in the previous subsection. In other words, only their thermal momentum changes. Initially it is given by  a thermal distribution at $T=$ 400 MeV and then cools down to 200 MeV through the scattering with the artificial partons, which are assumed to have a thermal distribution with  $T=$ 200 MeV.

In this case Eq.~(\ref{box2}) is still valid. The only difference is that  the initial momentum distributions of $p_1^*$ and $p_2^*$ correspond to  the thermal distribution of $T=$ 400 MeV. Then they gradually change with time, through scattering with the artificial partons, to the distribution corresponding to  $T=$ 200 MeV.
Simply expressed, though it is not quite true, Eq.~(\ref{boxi}) will be like
\begin{eqnarray}
P_\phi(t^\prime) = \mathcal{W}_p(T=400~{\rm MeV})\mathcal{W}_r\nonumber\\
+\mathcal{W}_p(T=399~{\rm MeV})\mathcal{W}_r- \mathcal{W}_p(T=400~{\rm MeV})\mathcal{W}_r\nonumber\\
+\mathcal{W}_p(T=398~{\rm MeV})\mathcal{W}_r- \mathcal{W}_p(T=399~{\rm MeV})\mathcal{W}_r\nonumber\\
...\nonumber\\
+\mathcal{W}_p(T=200~{\rm MeV})\mathcal{W}_r- \mathcal{W}_p(T=201~{\rm MeV})\mathcal{W}_r.
\end{eqnarray}
However, if the radius  or $\sigma$ in the Wigner function of Eq.~(\ref{1sWig}) depends on the temperature, Eqs.~(\ref{box1}) and (\ref{box2}) are not valid any more:
\begin{eqnarray}
\mathcal{W}_p(p_1^*, p_2^*,T; t_i+\varepsilon) \neq \mathcal{W}_p(p_1^*, p_2^*,T; t_{i+1}-\varepsilon),
\label{temp-ne}
\end{eqnarray}
because the temperature at $t=t_i+\varepsilon$ is different from that at $t=t_{i+1}-\varepsilon$.
Therefore, it is necessary to add the rate $\Gamma_{\rm local}$ as in Eqs.~(\ref{gamma}) and (\ref{joerg}), which is expressed by~\cite{Villar:2022sbv,Villar:2021loy}
\begin{eqnarray}
\Gamma_{\rm local}(t)={\rm Tr}\bigg(\frac{d\rho_\Phi}{d\sigma(T)}\frac{d\sigma(T)}{dt}\rho_N\bigg).
\label{pol}
\end{eqnarray}
Then Eq.~(\ref{boxi}) changes to 
\begin{eqnarray}
P_\phi(t^\prime) = \mathcal{W}_p(p_1^*, p_2^*,T; 0)\mathcal{W}_r\nonumber\\
+\mathcal{W}_r\int_0^{t_1}dt~(\partial \mathcal{W}_p(p_1^*, p_2^*; t)/\partial \sigma)(\partial \sigma/\partial t) \nonumber\\
+\mathcal{W}_p(p_1^*, p_2^*,T; t_1+\varepsilon)\mathcal{W}_r- \mathcal{W}_p(p_1^*, p_2^*,T; t_1-\varepsilon)\mathcal{W}_r\nonumber\\
+\mathcal{W}_r\int_{t_1}^{t_2}dt~(\partial \mathcal{W}_p(p_1^*, p_2^*; t)/\partial \sigma)(\partial \sigma/\partial t) \nonumber\\
+\mathcal{W}_p(p_1^*, p_2^*,T; t_2+\varepsilon)\mathcal{W}_r- \mathcal{W}_p(p_1^*, p_2^*,T; t_2-\varepsilon)\mathcal{W}_r\nonumber\\
...\nonumber\\
+\mathcal{W}_r\int^{t^\prime}dt~(\partial \mathcal{W}_p(p_1^*, p_2^*; t)/\partial \sigma)(\partial \sigma/\partial t) \nonumber\\
+\mathcal{W}_p(p_1^*, p_2^*,T; t^\prime+\varepsilon)\mathcal{W}_r- \mathcal{W}_p(p_1^*, p_2^*,T; t^\prime-\varepsilon)\mathcal{W}_r.
\label{boxTi}
\end{eqnarray}

Since nothing changes between $t=0$ and $t=t_1-\varepsilon$ except of the temperature dependent $\sigma$ 
\begin{eqnarray}
\int_0^{t_1}dt~(\partial \mathcal{W}_p(p_1^*, p_2^*; t)/\partial \sigma)(\partial \sigma/\partial t) \nonumber\\
=\mathcal{W}_p(p_1^*, p_2^*,T; t_1-\varepsilon)- \mathcal{W}_p(p_1^*, p_2^*,T; 0),
\label{cancel1}
\end{eqnarray}
ignoring the temperature change between $t=t_1-\varepsilon$ and $t=t_1$. Therefore, one arrives the same result as in Eq.~(\ref{boxf}):
\begin{eqnarray}
P_\phi(t^\prime) = \mathcal{W}_p(p_1^*, p_2^*,T; t^\prime+\varepsilon)\mathcal{W}_r.
\label{boxf2}
\end{eqnarray}

\begin{figure}[h]
\centerline{
\includegraphics[width=9 cm]{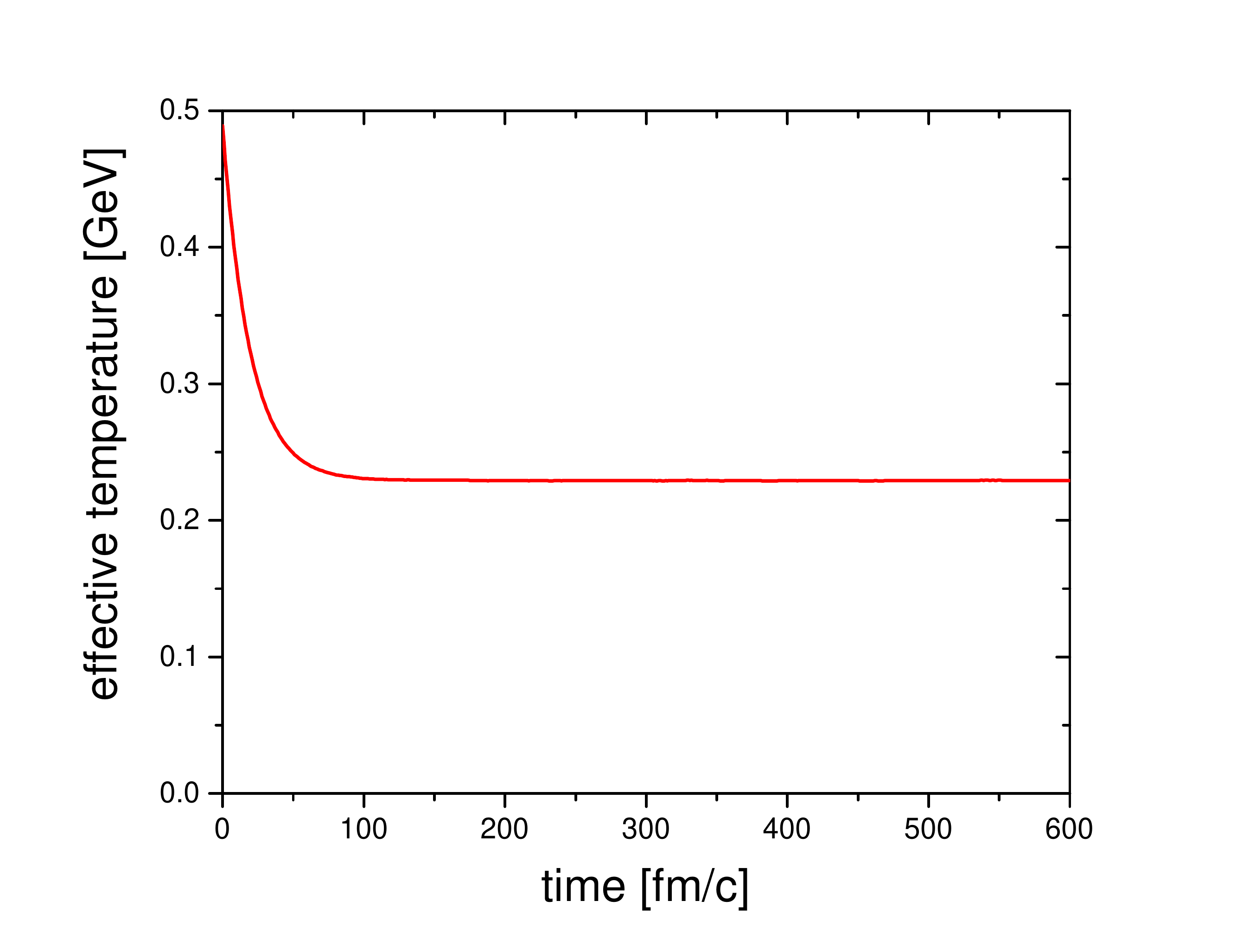}}
\centerline{
\includegraphics[width=9 cm]{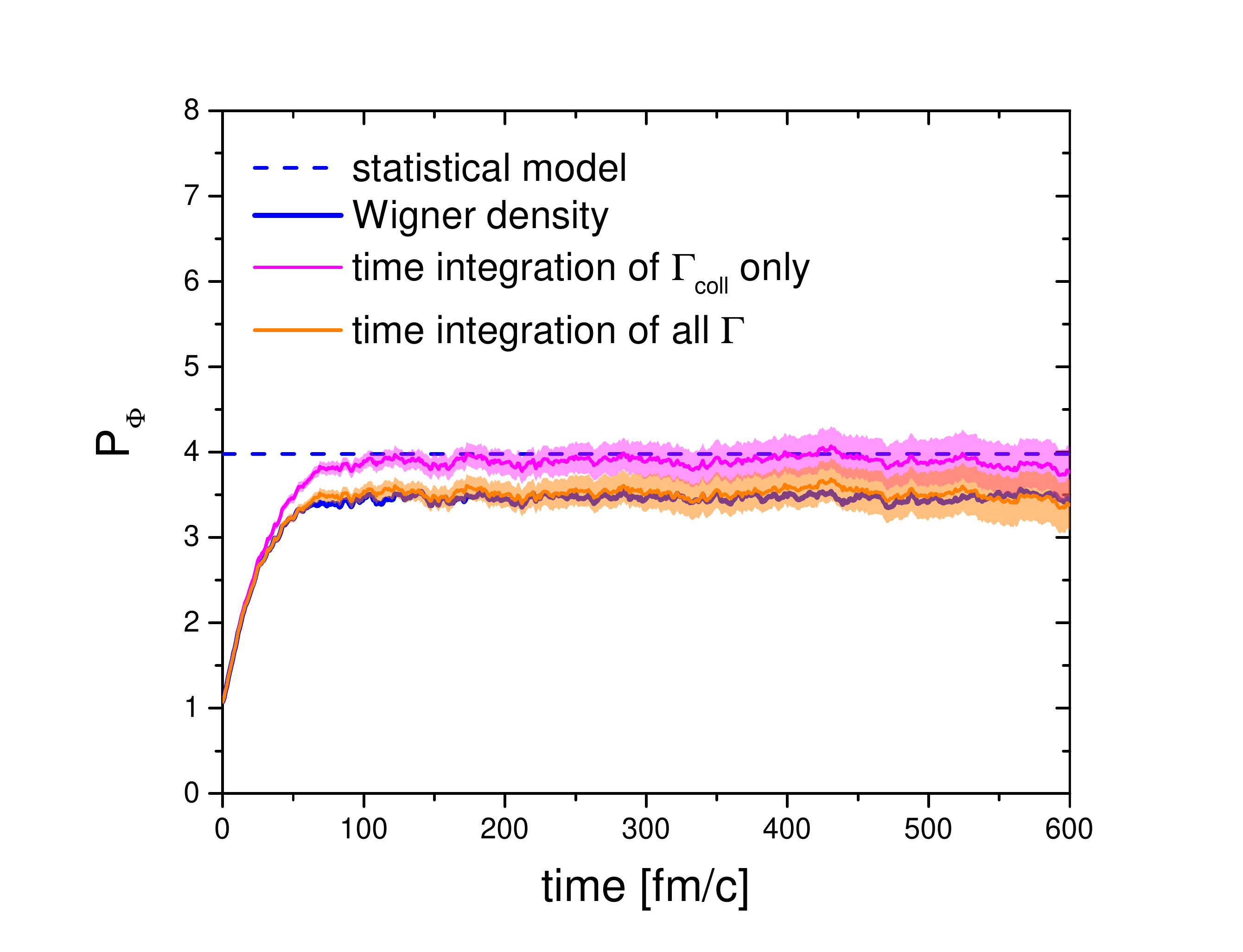}}
\caption{(Color online) (upper) Effective temperature of (anti)charm quarks as a function of time and (lower) the same as figure~\ref{box} but with and without the contributions from $\Gamma_{\rm local}$.}
\label{box-temp}
\end{figure}

The upper panel of Fig.~\ref{box-temp} shows the time evolution of the effective temperature, which is defined by 
\begin{eqnarray}
T_{\rm eff}=\frac{2}{3}\langle E_{kin} \rangle,
\label{effT}
\end{eqnarray}
where $\langle E_{kin} \rangle$ is the mean value of the  (anti)charm kinetic energy.
We note that  due to the approximation of Eq.~(\ref{effT}) the initial and the final effective temperatures are a bit higher than the real initial and final temperatures, which are, respectively, 400 MeV and 200 MeV.  One can see that the temperature reaches its final value before $t=$ 100 fm/c. The radius of quarkonium is simply modeled as
\begin{eqnarray}
\sqrt{\langle r^2 \rangle}=0.5 \bigg(\frac{T_{\rm eff}}{0.2~ {\rm GeV}}\bigg)^2~{\rm [fm]}
\end{eqnarray}
such that $\sqrt{\langle r^2 \rangle}=0.5$ fm at $T=$ 200 MeV, a reasonable approximation to the lattice results \cite{Lafferty:2019jpr}.
The lower panel of Fig.~\ref{box-temp} corresponds to  Fig.~\ref{box} . We have added the magenta line, which is the result if we apply only the collisional rate, $\Gamma_{\rm coll}$, whereas the orange line is obtained if we take the sum of  both rates, $\Gamma_{\rm local}$ and $\Gamma_{\rm coll}$.
The multiplicity of charmonia starts from a lower value than in equilibrium at $T=$ 200 MeV because a larger thermal momentum lowers the Wigner projection.
The orange line recovers the equilibrium multiplicity at about the same time when the temperature of the box reaches its final value.  One can clearly see that the inclusion of $\Gamma_{\rm local}$ is necessary to obtain a results consistent with the assumed equilibrium.

\subsection{expanding (anti)charm quarks}

\begin{figure}[h]
\centerline{
\includegraphics[width=9 cm]{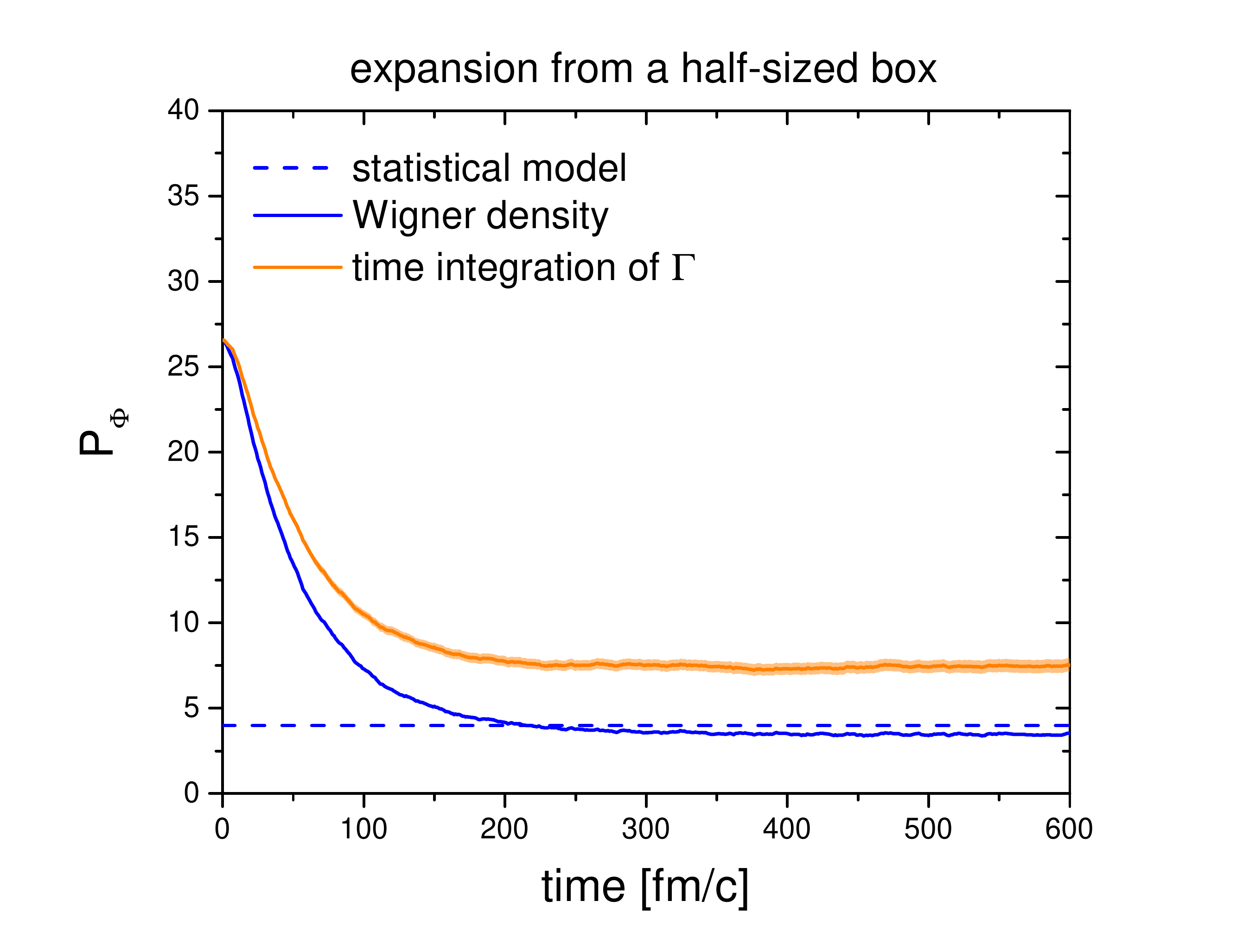}}
\caption{(Color online) Same as figure~\ref{box} but initial spatial distribution of (anti)charm quarks is restricted to a half-sized box.}
\label{box-expan}
\end{figure}

Now we confine the initial (anti)charm quarks within a smaller box of size $50^3~ {\rm fm^3}$ in the center of the large box, assuming a momentum distribution of the (anti)charm quarks, corresponding to $T=$ 200 MeV. The density of the charm and anticharm quarks in this smaller box is therefore 8 times higher as compared to the above discussed configuration.

As time passes the charm density decreases  and the number of charmonia converges to that expected for a system in equilibrium in the full volume. One can see in Fig.~\ref{box-expan} that the multiplicity, which is given by  the solid blue line, indeed converges to the number expected for a system in equilibrium. The time integration of $\Gamma_{\rm coll}$, however, does not catch up with the decrease and remains higher.
The reason is that $\mathcal{W}_r$ in Eq.~(\ref{boxi}) now depends on time and we obtain
\begin{eqnarray}
P_\phi(t^\prime) = \mathcal{W}_p(p_1^*, p_2^*; 0)\mathcal{W}_r(0)\nonumber\\
+\mathcal{W}_p(p_1^*, p_2^*; t_1+\varepsilon)\mathcal{W}_r(t_1+\varepsilon)~~~~~~~~~~~~\nonumber\\
- \mathcal{W}_p(p_1^*, p_2^*; t_1-\varepsilon)\mathcal{W}_r(t_1-\varepsilon)\nonumber\\
+\mathcal{W}_p(p_1^*, p_2^*; t_2+\varepsilon)\mathcal{W}_r(t_2+\varepsilon)~~~~~~~~~~~~\nonumber\\
- \mathcal{W}_p(p_1^*, p_2^*; t_2-\varepsilon)\mathcal{W}_r(t_2-\varepsilon)\nonumber\\
...\nonumber\\
+\mathcal{W}_p(p_1^*, p_2^*; t^\prime+\varepsilon)\mathcal{W}_r(t^\prime+\varepsilon)~~~~~~~~~~~~\nonumber\\
- \mathcal{W}_p(p_1^*, p_2^*; t^\prime-\varepsilon)\mathcal{W}_r(t^\prime-\varepsilon),
\end{eqnarray}
Eqs.~(\ref{box1}) and (\ref{box2}) are still valid, but the two terms do not cancel any longer:
\begin{eqnarray}
\mathcal{W}_p(p_1^*, p_2^*; t_i+\varepsilon)\mathcal{W}_r(t_i+\varepsilon)\nonumber\\
 \ne \mathcal{W}_p(p_1^*, p_2^*; t_{i+1}-\varepsilon)\mathcal{W}_r( t_{i+1}-\varepsilon),
\label{problem}
\end{eqnarray}
because $\mathcal{W}_r(t_i+\varepsilon)$ is larger than $\mathcal{W}_r( t_{i+1}-\varepsilon)$ due to the spatial diffusion of (anti)charm quarks. This is why the equilibrium distribution of Eq.~(\ref{boxf}) cannot be achieved in this scenario.


\subsection{cooling and expanding (anti)charm quarks}\label{sim-HIC}

\begin{figure}[h]
\centerline{
\includegraphics[width=9 cm]{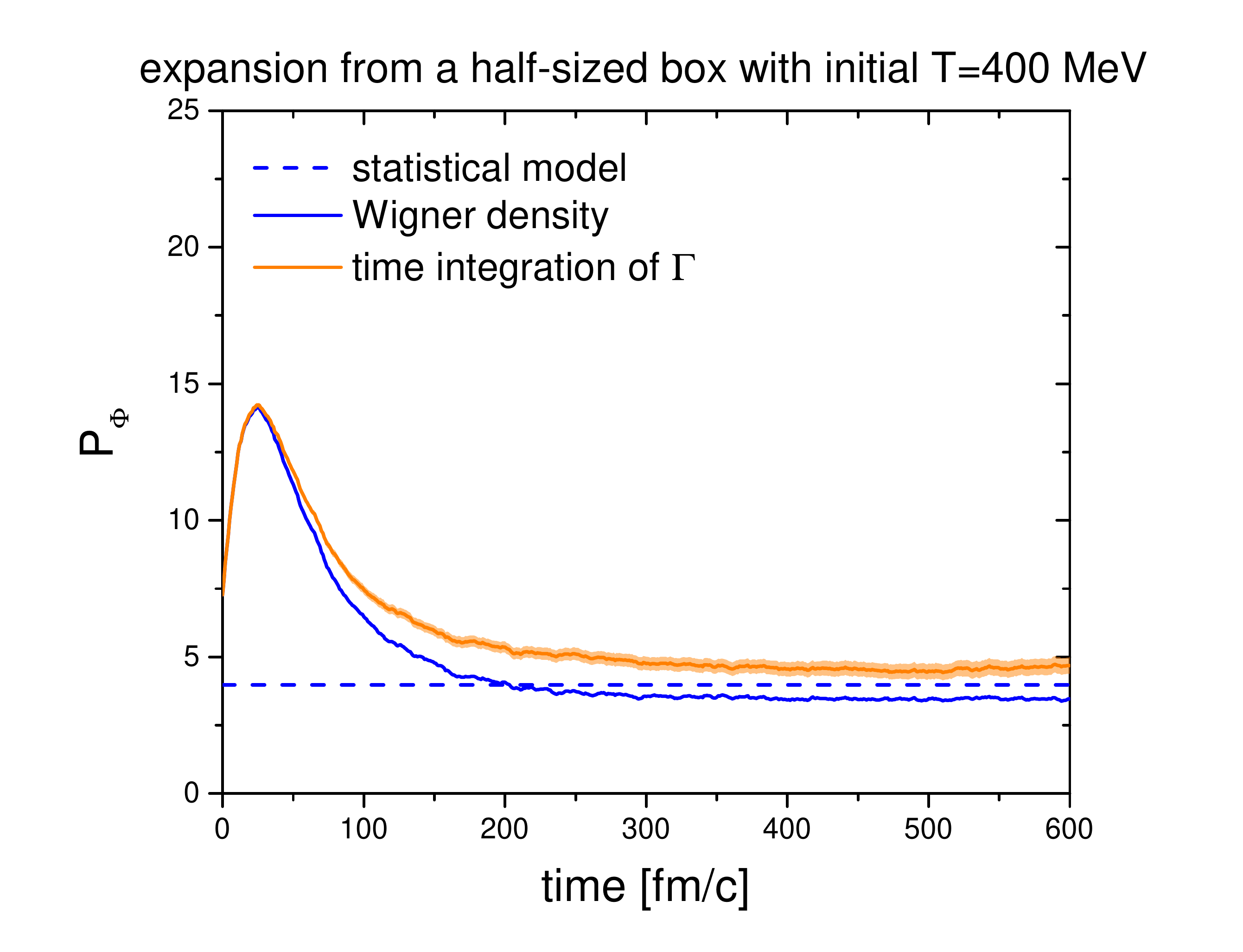}}
\caption{(Color online) Combination of figure~\ref{box-temp} and figure~\ref{box-expan} for the initial conditions of (anti)charm quarks }
\label{box-HIC}
\end{figure}

Now we imitate heavy-ion collisions by combining the  two previous scenarios, in other words, heavy (anti)quarks are initially at a high temperature and densely populated in a small volume. As before the initial temperature is given by 400 MeV and the initial volume by $50^3~ {\rm fm^3}$. Then they cool down and expand in space.

Therefore the projection probability first increases with time due to the momentum loss, as in Fig.~\ref{box-temp}, and then decreases due to the spatial diffusion, as in Fig.~\ref{box-expan}.
However, one can see in Fig.~\ref{box-HIC} the multiplicity, obtained by the time integration of $\Gamma_{\rm coll}$,  differs from that obtained by applying the Wigner projection directly and from that, which is given by the statistical model.

\section{spatial diffusion term}\label{diff-sec}

The discrepancies between the statistical model and the time integration of $\Gamma$ in Figs.~\ref{box-expan} and \ref{box-HIC} originate from the calculations of $d\rho_N/dt$. 
Since $\rho_N$ is the density operator of $N$ (anti)charm quarks, it includes both spatial and momentum information.
But only momentum space information has been taken from the comparison between Eq.~(\ref{potential}) and Eq.~(\ref{impulse}).
In other words, only the interaction terms are taken into account and the kinetic terms (the free streaming) are neglected in Eq.~(\ref{impulse}). The spatial diffusion is attributed to the kinetic terms in $K_1$ and $K_2$ of Eq.~(\ref{hamiltonian}). In principle, it must not contribute to $d\rho_N/dt$ because heavy quark and heavy antiquak are bound by the potential $V_{12}$ and move together as shown in Eq.~(\ref{h12}).  However,  in standard  cascade simulations we have only free propagation and instant scatterings~\cite{Gyulassy:1982pe} and it is still challenging to properly implement microscopic potentials in numerical simulations~\cite{Villar:2022sbv,Krenz:2018joj}.

We therefore propose that the following term should be added to Eq.~(\ref{gamma}) if a standard cascade approach is employed
\begin{eqnarray}
\Gamma_{\rm diff}(t)={\rm Tr}\bigg(\frac{d\rho_\Phi}{d\vec{r}}\cdot\frac{d\vec{r}}{dt}\rho_N\bigg),
\label{diff}
\end{eqnarray}
and for an S-state it will be (see Eq.~(\ref{1sWig})) of the form
\begin{eqnarray}
\Gamma_{\rm diff}(t)\sim -\frac{2}{\sigma^2}\vec{r}\cdot \vec{v} ~W_S(r,p),
\label{diff-s}
\end{eqnarray}
where $\vec{v}=d\vec{r}/dt$ with $\vec{r}$ being $\vec{r}_Q-\vec{r}_{\bar{Q}}$ in their center-of-mass frame.
Then the inequality of Eq.~(\ref{problem}) can be removed through
\begin{eqnarray}
\mathcal{W}_p(p_1^*, p_2^*; t_i+\varepsilon)\mathcal{W}_r(t_i+\varepsilon)\nonumber\\
-\mathcal{W}_p(p_1^*, p_2^*; t_{i+1}-\varepsilon)\mathcal{W}_r( t_{i+1}-\varepsilon)\nonumber\\
+\int_{t_i}^{t_{i+1}}dt~\Gamma_{\rm diff}=0.
\label{cancel2}
\end{eqnarray}

In fact, if the bound state is perfectly described in the simulations, even $\Gamma_{\rm local}$ is not needed, because the equality of Eq.~(\ref{temp-ne}) will dynamically be restored:
\begin{eqnarray}
\mathcal{W}_p(p_1^*, p_2^*,T; t_i+\varepsilon) = \mathcal{W}_p(p_1^*, p_2^*,T; t_{i+1}-\varepsilon).
\end{eqnarray}

As temperature decreases, the binding of quarkonium will be stronger so that the relative distance $r$ decreases and the relative momentum $p$ increases, which compensates the change of $\sigma$ with temperature.

\begin{figure}[h]
\centerline{
\includegraphics[width=9 cm]{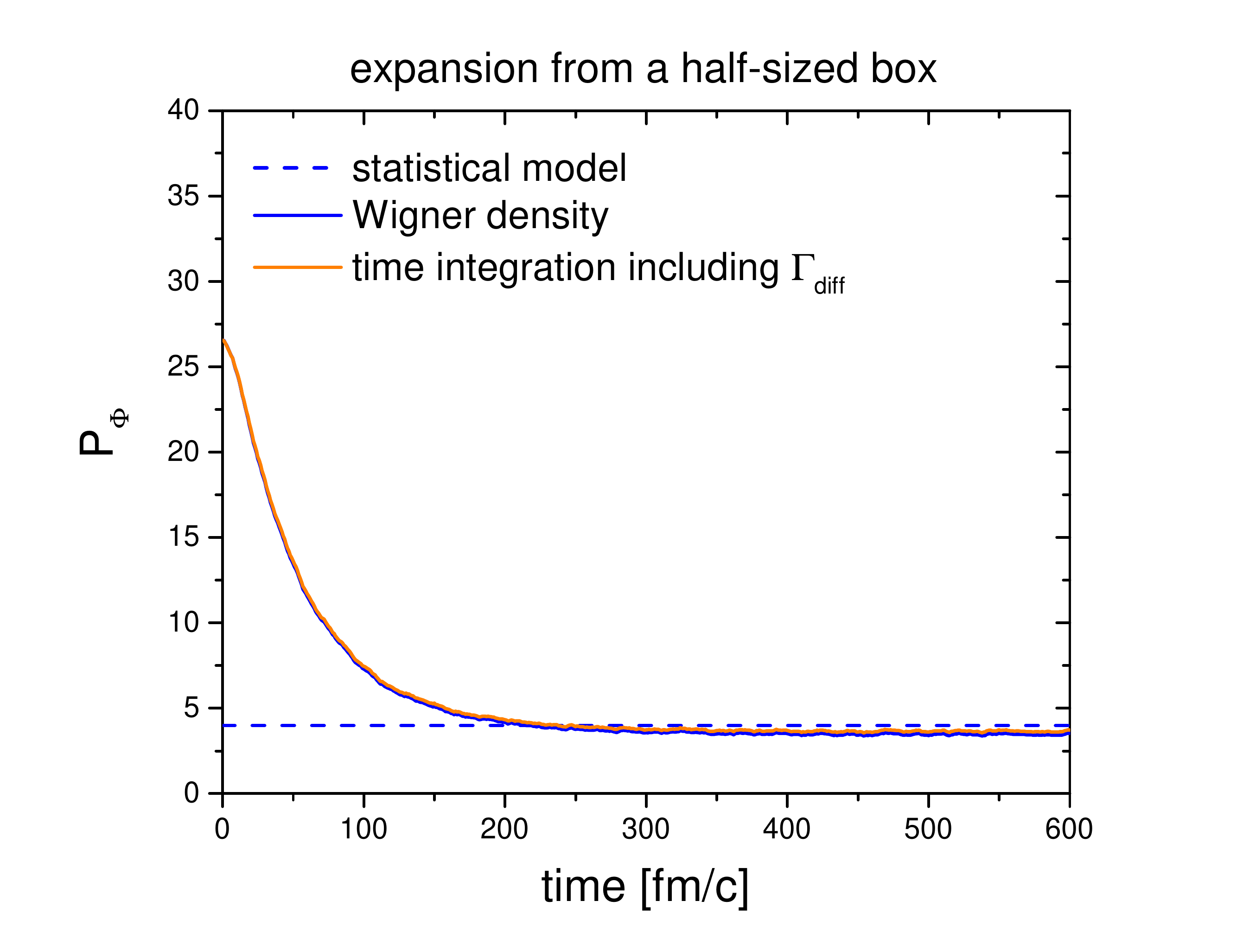}}
\centerline{
\includegraphics[width=9 cm]{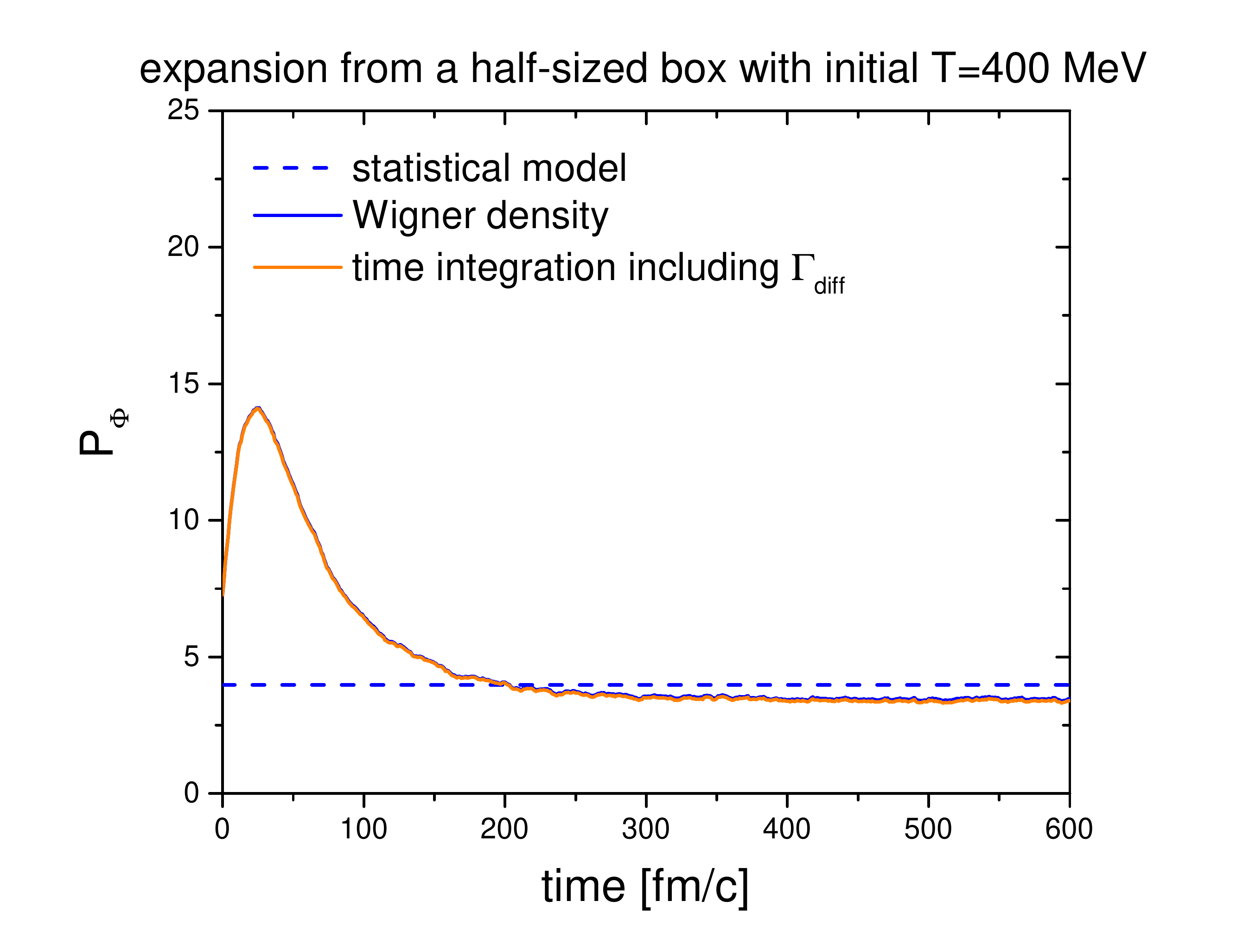}}
\caption{(Color online) Same as figures~\ref{box-expan} and \ref{box-HIC} but including $\Gamma_{\rm diff}$ of Eq.~(\ref{diff-s})}
\label{boxnew}
\end{figure}

Figure~\ref{boxnew} shows the time evolution of the multiplicity of $\Phi$ 
including $\Gamma_{\rm diff}$ in comparison the cases already studied in figures~\ref{box-expan} and \ref{box-HIC}. One finds that including $\Gamma_{diff}$  the time integration of $\Gamma$ is now in good agreement with the statistical model predictions.
Considering the results of our study of the $\Phi$ multiplicity in a box, the best and simplest method to obtain the asymptotically correct values is to add to  $\Gamma$ a diffusion rate $\Gamma_{\rm diff}$
\begin{eqnarray}
\Gamma(t)=\Gamma_{\rm local}(t)+\Gamma_{\rm coll}(t)+\Gamma_{\rm diff}(t)\nonumber\\
= \sum_{i=1,2}\sum_{j\ge 3}\sum_\nu \int d^3r_1d^3p_1 ... d^3r_N d^3p_N(2\pi)^{3N}\nonumber\\
\times\rho_\Phi(r_1,p_1;r_2,p_2)
\bigg\{\delta\bigg(t-t_{ij}(\nu)\bigg)\rho_N(t+\varepsilon)\nonumber\\
-\delta\bigg(t-t_{ij}(\nu-1)\bigg)\rho_N(t+\varepsilon)\bigg\},~~~
\label{new}
\end{eqnarray}
where $\nu$ means $\nu$ th scattering of $i=1$ or of $i=2$. 

Consequently, to be consistent with statistical model predictions, one has to supplement the old projection probability of particles $i=1,2$ (the second term in the bracket)  by a new one (the first term in the bracket) whenever a scattering happens to $i=1$ or to $i=2$.  Then the change of the temperature,  reflected in the change of $\sigma$, and of the spatial separation of (anti)charm quarks between $t=t_{ij}(\nu-1)$ and $t=t_{ij}(\nu)$ will completely be canceled.  This approach is also more natural than Eq.~(\ref{gam-old}), which assumes an instant interaction between $t-\varepsilon$ and $t+\varepsilon$,  because it increases $\varepsilon$ to the time between two scatterings. We note that the combination of Eq.~(\ref{cancel1}) and Eq.~(\ref{cancel2}) yields
\begin{eqnarray}
\mathcal{W}_p(p_1^*, p_2^*; t_i+\varepsilon)\mathcal{W}_r(t_i+\varepsilon)\nonumber\\
-\mathcal{W}_p(p_1^*, p_2^*; t_{i+1}-\varepsilon)\mathcal{W}_r( t_{i+1}-\varepsilon)\nonumber\\
+\mathcal{W}_r(t_i)\int_{t_i}^{t_{i+1}}dt~\frac{\partial \mathcal{W}_p(p_1^*, p_2^*; t)}{\partial \sigma}\frac{\partial \sigma}{\partial t} \nonumber\\
+\mathcal{W}_p(p_1^*, p_2^*; t_i)\int_{t_i}^{t_{i+1}}dt~\frac{\partial \mathcal{W}_r(t)}{\partial \vec{r}}\cdot\frac{\partial \vec{r}}{\partial t} =0.
\end{eqnarray}

\section{summary}\label{summary}

The Remler formalism has been advanced to study the production of  $J/\psi$  in a thermalized expanding system. In this study we have tested the Remler approach for $J/\psi$ production in thermalized and thermalizing boxes, composed of c and $\bar c$ quarks.
The goal was to verify whether the numerical, Monte Carlo based, realization of the Remler algorithm gives the right asymptotic solution for $t\to\infty$ ,  which can be calculated analytically.

We started out with calculations  in a completely thermalized box, where we find  that the Remler formula produces the results which are consistent with that of a statistical model calculation. Then three different types of thermalizing boxes have been investigated.

In the first scenario the initial temperature of the charm quarks is high. They cool down with time through collisions with background particles and finally reach thermal equilibrium at a lower temperature. The multiplicity of the charm quarks is kept constant.
We have found that the temperature derivative of the Wigner function is required to assure that the Remler approach agrees  for large times with statistical model calculations.

In the second scenario charm and anticharm quarks are initially concentrated in a smaller box and then diffuse in space.  The last scenario, which  we studied, is the combination of the first and second one: The initial temperature of the charm quark is higher than that of the background particles and (anti-)charm quarks are concentrated initially in the central region of the box. Then they cool down and diffuse in space. This is  a  simple model for the expansion of the midrapidity QGP, which is created in  heavy-ion collisions.
We have found that in the second and third scenarios the discrepancy between the multiplicities, calculated in the Remler approach and in a statistical model, does not disappear even for $t\to\infty$.

We  identified the origin of this discrepancy: it is caused  by the fact that in the numerical realization of the Remler algorithm, as presented in~\cite{Gyulassy:1982pe} for deuterons, the expansion of the system between two subsequent collisions is not taken properly into account. Neglecting the potential, it treats the c and $\bar c$ quarks as freely moving particles between two collisions whereas in reality they are bound when they form a quarkonium. Introducing a diffusion rate, which adds to the local rate and the collision rate, this discrepancy disappears.
We note that recently also an approach has been advanced which includes the $c\bar c$ potential in an approximate way \cite{Villar:2022sbv}. It would be interesting to verify whether there equilibrium is obtained without a spacial diffusion rate. 
  In conclusion, we have found that also for an expanding  system, which cools down, the Monte Carlo realization of the Remler formalism describes correctly the approach to equilibrium .

\section*{Acknowledgements}
The authors acknowledge valuable discussions with P.-B. Gossiaux. We acknowledge support by the Deutsche Forschungsgemeinschaft (DFG, German Research Foundation) through the grant CRC-TR 211 'Strong-interaction matter under extreme conditions' - Project number 315477589 - TRR 211. 
This work is supported by the European Union’s Horizon 2020 research and innovation program under grant agreement No 824093 (STRONG-2020).
The computational resources have been provided by the LOEWE-Center for Scientific Computing and the "Green Cube" at GSI, Darmstadt and by the Center for Scientific Computing (CSC) of the Goethe University.

\bibliography{heat-bath}

\end{document}